\documentstyle[prb,aps,multicol,epsf]{revtex}

\begin{document}
\draft

\title{Spin wave dispersion softening in the \protect\\ ferromagnetic Kondo
 lattice model for manganites}
\author{F.Mancini$^{a}$, N.B. Perkins$^{a,b}$, and N.M. Plakida$^{b}$}
\address{
$^{a}$ Dipartimento di Scienze Fisiche ``E.R.Caianiello'' --
Unit\`a INFM di Salerno\protect\\, Universit\`a degli Studi di
Salerno, 84081 Baronissi (SA), Italy\\
$^{b}$Joint Institute for Nuclear Research, BLTP, Dubna, Moscow
region, 141980, Russia }

\maketitle
\begin{abstract}\widetext
Spin dynamics is calculated in the ferromagnetic (FM) state of the
generalized Kondo lattice model taking into account strong on-site
correlations between $e_g$ electrons and antiferromagnetic (AFM)
exchange among $t_{2g}$ spins.
 Our study suggests that competing FM double-exchange
and AFM super-exchange interaction lead to a rather nontrivial
spin-wave spectrum. While spin excitations have a conventional
$Dq^2$ spectrum in the long-wavelength limit, there is a strong
deviation from the spin-wave spectrum of the isotropic Heisenberg
model close to  the zone boundary. The relevance of our results to
the experimental data are  discussed.
\end{abstract}

\begin{multicols}{2}

\narrowtext

The revival in the study of manganites has led to experimental
re-examination of their different properties and interplay between
nontrivial magnetism and highly anisotropic charge transport. One
of the puzzling features  is the non-universality of the magnetic
and transport properties of FM manganites at doping $x\sim 0.3$
\cite{173}. According to the conventional theory of the Double
Exchange (DE), the spin dynamics of the FM state that evolves at
temperatures below the Curie temperature $T_C$ is expected to be
of nearest-neighbor Heisenberg type. This picture seems to be
indeed reasonably accurate for manganites with high value of $T_C$
\cite{hal6}. Recently, however, several  experimental results have
shown a  strong deviation of the spin-wave dispersion (SWD) from
the typical Heisenberg behavior. The unexpected softening of the
SWD at the zone boundary has been observed in several manganites
with low but very different $T_C$: Pr$_{0.63}$Sr$_{0.37}$MnO$_3$
($T_C$=301 K) \cite{1}, La$_{0.7}$Ca$_{0.3}$MnO$_{3}$  ($T_C$=242
K) \cite{2} and Nd$_{0.7}$Sr$_{0.3}$MnO$_3$  ($T_C$=198 K)
\cite{3}. These observations are very important as they indicate
that some aspects of spin dynamics in manganites have not been
entirely understood yet. During last few years this problem was
discussed by several theoreticians. Khaliullin and Kilian in
Ref.\onlinecite{4} proposed a theory of anomalous softening in FM
manganites based on the modulation of magnetic exchange bonds by
orbital degree of freedom of double-degenerate $e_g$ electrons.
They found out that charge and coupled orbital-lattice
fluctuations can be considered as the main origin of the softening
phenomena. Soloviev and Terakura in Ref.\onlinecite{5} have argued
that the softening of spin wave at the zone boundary and the
increase of the spin-wave stiffness constant with doping have
purely magnetic origin.

The DE mechanism as an explanation of ferromagnetic order in doped
manganites \cite{zen} was proposed shortly after the first
experiments in these compounds \cite{Jon}. In this model it is
assumed that the hopping of $e_g$ electrons between neighboring
sites is easier if the local spin on the sites are parallel, so an
effective ferromagnetic coupling between the local spins is
induced by the conduction electrons lowering their kinetic energy.
For quite a long time it was believed that DE model gives full
understanding of the FM metallic state in manganites. Later it was
shown both from theory and experiment that, in fact, it provides
only a qualitative explanations and it is necessary to extend DE
model by taking into account the orbital degree of freedom of
double-degenerate $e_g$ electrons, phonons, AFM superexchange
interaction between localized spins , etc. Particularly, for the
Curie temperature the value estimated by pure DE was much larger
than the experimentally observed one. Taking into account AFM
superexchange interaction among localized spins of $t_{2g}$
electrons gives possibility to diminish the value of $T_C$ in the
framework of DE model \cite{dzero}. Recently, a Monte Carlo study
of the Ferromagnetic Kondo Lattice Model (FKLM) for doped
manganites (see, Ref.\onlinecite{YI}) confirmed the scaling of the
$T_C$ suppression  with a bandwidwidth narrowing induced by the
antiferromagnetic frustration. In our recent paper \cite{jac} we
have also shown that AFM superexchange interaction plays very
important role in determing the phase diagram in manganites.

The aim of our work is to show that the competition between AFM
superexchange interaction among localized spins and
double-exchange interaction among itinerant electrons is one of
the  mechanisms responsible for the spin wave softening phenomena.

In the present paper we continue to study the spin dynamics within
the generalized one-orbital FMKL \cite{6}. The effective
Hamiltonian of FMKL model \cite{kubo,dag} can be written as
$$H=\sum_{ij}(t_{ij}-\mu \delta_{ij})c^{\dagger}(i)c(j)+U\sum_{i}
n_{\uparrow}(i)n_{\downarrow}(i)$$
\begin{equation}
-J_H\sum_{i}{\bf S}(i){\bf s}(i) + 1/2\sum_{ij}J_{ij}{\bf
S}(i){\bf S}(j) ~, \label{h1}
\end{equation}
where $c(i)$ and $c^{\dagger}(i)$ are annihilation and creation
operators for electrons in the spinor notation at site ${\bf
R}(i)$; $$ c(i)=\left (
\begin{array}{c}
c_{\uparrow}(i)\\
c_{\downarrow}(i)
\end{array}
\right )~, $$ $\mu$ is the chemical potential;
$n_{\sigma}(i)=c_{\sigma}^{\dagger}(i)c_{\sigma}(i)$ is the
density operator of electrons with spin $\sigma$; the spin
operator  is given by $ s_k(i)=\frac{1}{2}c^{\dagger}(i)\sigma_k
c(i)$, $\sigma_k$ ($k=1,2,3$) are the Pauli matrices; ${\bf S}(i)$
is the localized Mn core spin of $t_{2g}$ electrons with $S=3/2 $;
$t_{ij}$ is the nearest-neighbor hopping parameter; $J_H$ is the
FM Hund coupling interaction among the localized and delocalized
spin subsystems, $J_{ij}$ is the nearest-neighbor
antiferromagnetic superexchange interaction among localized spins.

The magnetic properties of the above given Hamiltonian has a dual
character: there exists a purely Heisenberg-like contribution from
the exchange interaction among localized spins and the itinerant
contribution driven by electronic excitations. The itinerant
contribution leads to the effective FM exchange interaction
between localized spins due to the DE mechanism. To treat
appropriately this dual character of the magnetism one should
describe both contributions on the equal level of approximation as
well as retain SU(2) spin symmetry of the Hamiltonian. To this
end, we employ the Composite operator method (COM) approach
\cite{8} by introducing  a new set of operators which well
describe collective excitations in the system and are constructed
from the original electron operators. These composite operators
can have fermionic or bosonic character. They are created by
interactions among the electrons and the localized spins, and,
therefore, their properties will be determined by the dynamics and
boundary conditions and must be computed self-consistently.

First we start by discussing the electronic excitations of the
system, responsible for the itinerant contribution to spin
dynamics. We introduce  the 4-component fermionic basis:
\begin{eqnarray}
\psi (i)= \left(
\begin{array}{c}
\xi_{\uparrow} (i)\\
\eta_{\uparrow} (i)\\
\xi_{\downarrow} (i)\\
\eta_{\downarrow} (i)\\
\end{array}
\right )~, \label{psi}
\end{eqnarray}
where $\xi_{\sigma}(i)$ and $\eta_{\sigma}(i)$ are Hubbard
composite excitations:
\begin{eqnarray}
\begin{array}{c}
\xi_{\sigma} (i)=c_{\sigma}(i) (1-n_{\bar\sigma}(i))\\
\eta_{\sigma} (i)=c_{\sigma}(i) n_{\bar\sigma}(i)~~.
\end{array}
\label{hub}
\end{eqnarray}
The $\eta$ excitation describes an electron restricted to move on
sites already occupied with an electron of opposite spin whereas
$\xi$ demands that there be no prior occupancy on the site.

Let us consider the equation of  motion for the field $\psi (i)$:
\begin{eqnarray}
\imath \frac{\partial}{\partial t} \psi (i)=j(i)=[\psi(i),H]~,
\label{j}
\end{eqnarray}
where $j(i)$ is the current operator. By considering the
Hamiltonian given by (\ref{h1}) , we obtain the following
expression for the current operator:

\end{multicols}

\widetext

\begin{eqnarray}
j(i)= \left(
\begin{array}{c}
-\mu\xi_{\uparrow} (i)-6tc_{\uparrow}^{\alpha}(i)-
6t\pi_{\uparrow}(i)-J_H/2~[\xi_{\downarrow}(i)S^-(i)+
\xi_{\uparrow}(i)S^z(i)] \\[1mm]
-(\mu-U)\eta_{\uparrow} (i)+6t\pi_{\uparrow}(i)-J_H/2~
[\eta_{\downarrow}(i)S^-(i)+\eta_{\uparrow}(i)S^z(i)]\\[1mm]
-\mu\xi_{\downarrow} (i)-6tc_{\downarrow}^{\alpha}(i)-
6t\pi_{\downarrow}(i)-J_H/2~[\xi_{\uparrow}(i)S^+(i)-
\xi_{\downarrow}(i)S^z(i)] \\[1mm]
-(\mu-U)\eta_{\downarrow} (i)+6t\pi_{\downarrow}(i)-J_H/2~
[\eta_{\uparrow}(i)S^+(i)-\eta_{\downarrow}(i)S^z(i)]
\end{array}
\right)~, \label{eq}
\end{eqnarray}

\begin{multicols}{2}

\narrowtext

\noindent where $\pi
(i)=\frac{1}{2}\sigma^{\mu}n_{\mu}(i)c^{\alpha}(i)+ c(i)c^{\alpha
\dagger}c(i)$ is a higher order composite field;
$n_{\mu}(i)=c^{\dagger}(i)\sigma_{\mu}c(i)$ - the number ($\mu=0$)
and spin ($\mu=1,2,3$) density operator with the notation
$\sigma_{\mu}= ({\bf 1},\vec \sigma)$, $\sigma^{\mu}= (-{\bf
1},\vec \sigma)$, $\vec\sigma$ being the Pauli matrices; and the
notation $c^{\alpha}(i)=\sum_{j} \alpha (i,j) c(j)$ stands to
indicate the field $c$ on the first neighbor sites with
$F.T.\alpha (i,j)= \frac{1}{3}(\cos k_x+ \cos k_y +\cos k_z)$
(abbreviation $F.T.$ stands for Fourier transform).

Formally, we can write the $n$th element of the current as
\begin{eqnarray}
j_n(i)= \sum_j\sum_m\varepsilon_{nm}(i,j)\psi_m (j)+\delta
j_n(i)~. \label{jn}
\end{eqnarray}
By projecting out the correction part  $\delta j_n (i)$, we
linearize the Heisenberg equation of motion as
\begin{eqnarray}
\imath \frac{\partial}{\partial t} \psi_n (i)=
\sum_j\varepsilon_{nm}(i,j)\psi_m (j) ~, \label{lin}
\end{eqnarray}
where the energy matrix $\varepsilon (i,j)$ is  calculated by
means of the equation
\begin{equation}
\varepsilon (i,j)=\sum_l m(i,l) I^{-1}(l,j)~. \label{en1}
\end{equation}
To shorthand the notation we defined  the normalization matrix
$I(i,j)$ and  the frequency matrix $m(i,j)$: $$I(i,j)=<\{\psi
(i),\psi^{\dagger}(j)\}_{E.T.}>~,$$ $$m(i,j)=<\{j
(i),\psi^{\dagger}(j)\}_{E.T.}>~,$$ where $E.T.$ indicates the
anticommutator evaluated at  equal-time; the bracket $<...>$
denotes the thermal average on the grand canonical ensemble.

Straightforward calculations give us expressions for the Fourier
transforms of the matrices $I$ and  $m$. The normalization matrix
is diagonal and has the following matrix elements:
\begin{eqnarray}
\begin{array}{l}
I_{11}=1-<n_{\downarrow}>=1-n/2+<s_z>~,\\[1mm]
I_{22}=<n_{\downarrow}>=n/2-<s_z>~,\\[1mm]
I_{33}=1-<n_{\uparrow}>=1-n/2-<s_z>~,\\[1mm]
I_{44}=<n_{\uparrow}>=n/2+<s_z> ~;
\end{array}
\label{i}
\end{eqnarray}
where $n\equiv <n>$ is the particle density and $<s_z>=1/2<n_3>$.
The $m$ matrix is hermitian and the elements different from zero
are given by
\begin{eqnarray}\nonumber
m_{11}&=&-\mu I_{11}-6t\Delta_{\downarrow}-6t\alpha ({\bf
k})(I_{11}-I_{22}+
p_{\uparrow})\\
\nonumber
&-& J_H(a+b_2)~,\\
\nonumber
m_{12}&=&6t\Delta_{\downarrow}+6t\alpha ({\bf k})(-I_{22}+p_{\uparrow})~,\\
\nonumber m_{22}& =& -(\mu-U)I_{22}-6t\Delta_{\downarrow}-6t\alpha
({\bf k})p_{\uparrow}
+J_Hb_2~,\\
\nonumber m_{33}&=&-\mu I_{33}-6t\Delta_{\uparrow}-6t\alpha ({\bf
k})(I_{33}
- I_{44}+p_{\downarrow})\\
\nonumber
&-& J_H(-a+b_1)~,\\
\nonumber
m_{34}&=& 6t\Delta_{\uparrow}+6t\alpha ({\bf k})(-I_{44}+p_{\downarrow})~,\\
\nonumber m_{44}&=& -(\mu-U)I_{44}-6t\Delta_{\uparrow}-6t\alpha
({\bf k})p_{\downarrow}
+J_H b_1~,\\
\label{m}
\end{eqnarray}
where $\sigma=\uparrow (1)$ or $\sigma=\downarrow (2)$. Matrices
$m({\bf k})$ and $I$ involve correlation functions which we
defined as
\begin{eqnarray}
\begin{array}{ll}
\Delta_{\sigma}=&<\xi_{\sigma}^{\alpha}(i)\xi_{\sigma}^{\dagger}(i)>-
<\eta_{\sigma}^{\alpha}(i)\eta_{\sigma}^{\dagger}(i)>~,\\[1mm]
p_{\sigma}=&\frac{  1}{  4} [<n_{\mu}^{\alpha}(i)n_{\mu}(i)>+
2(-1)^{\sigma}<n^{\alpha}(i)n_{3}(i)>]\\[1mm]
&-<c_1(i)c_2(i)
[c^{\dagger}_2(i)c^{\dagger}_1(i)]^{\alpha}>~,\\[1mm]
a=&1/2<S_z>~,\\[1mm]
b_{\sigma}=&1/2
(<s^-(i)S^+(i)>-(-1)^{\sigma}<n_{\sigma}(i)S^z(i)>)~.
\end{array}
\label{def}
\end{eqnarray}
The parameters $\Delta_{\sigma}$ and $p_{\sigma}$ are static
inter-site correlation  functions which describe, respectively, a
constant shift of the bands and a bandwidth renormalization in
itinerant subsystem.

The retarded Green function (GF) can be written as $$
S^F(i,j)=<R\{\psi(i),\psi^{\dagger}(j)\}>= $$
\begin{equation}
\frac{\imath\Omega}{(2\pi)^4}\int d^3 k \int \omega \exp^{i{\bf
k}({\bf r}_i-{\bf r}_j)-i\omega(t_i-t_j)} S^F({\bf k},\omega)~,
\label{gfun}
\end{equation}
where the k-integration is over the Brillouin zone (BZ) and
$\Omega$ is the inverse volume of the BZ. By means of the
linearized Heisenberg equation (7) the Green's function has the
following expression: $$ S^F({\bf k}, \omega)=\frac{1}{\omega
-m({\bf k})I^{-1}({\bf k})}I({\bf k})~. $$ or, in  the spectral
form:
\begin{eqnarray}
S^F({\bf k},\omega)=\sum_{i=1}^{4}\frac{\sigma^i({\bf k})}
{\omega-E_i({\bf k})+i\delta} ~, \label{gff}
\end{eqnarray}
where the  energy spectra $E_i({\bf k})$ and the spectral
functions $\sigma^i({\bf k})$ can be easily evaluated. The energy
spectra $E_i({\bf k})$ are the eigenvalues of the matrix
$\varepsilon ({\bf k})$. Since they depend on a set of external
parameters such as electron density $n$, temperature $T$, Coulomb
interaction $U$, Hund coupling $J_H$ and AFM Heisenberg exchange
interaction $J$, and a set of internal parameters ($\mu, m,
\Delta_{\sigma}, p_{\sigma}, a, b_1, b_2$), they must be
calculated in a self-consistent way. Internal parameters are
expressed as expectation values of composite fields. If these
composite fields belong to the fermionic or bosonic basis they can
be expressed in terms of the corresponding GF. However, it may
happen that some of the parameters are expressed as expectation
values of higher-order composite fields that do not belong to the
basic set and we need to evaluate by imposing some symmetry
requirements, as it will be discussed later.

As we have already mentioned the spin dynamics of the present
model is governed by both localized and itinerant spin subsystems.
To treat these two contributions at the same level of the
approximation we consider a bosonic sector by defining a basis of
composite fields as
\begin{eqnarray}
B(i)= \left(
\begin{array}{c}
s^+ (i)\\
S^+ (i)
\end{array}
\right)~. \label{bos}
\end{eqnarray}
The procedure then is similar to the fermionic sector; the
propagator for the boson field $B(i)$ will contain another set of
parameters which must be self-consistently determined. After some
algebra the following expressions of the bosonic normalization
matrix $$\tilde I({\bf k})=F.T.<[B(i),B^{\dagger}(j)]_{E.T}>$$ and
the frequency matrix which is  given by the commutator of the
current and the bosonic fields $$\tilde m({\bf k})=F.T.<[\imath
\frac{\partial}{\partial t_i}B(i),B^{\dagger}(j)]>_{E.T.}$$ can be
easily derived:
\begin{eqnarray}
\begin{array}{l}
\tilde I_{11}=2<s_z>~,\\[1mm]
\tilde I_{12}=\tilde I_{21}=0~,\\[1mm]
\tilde I_{22}=2<S_z>~;
\end{array}
\label{tildei}
\end{eqnarray}

\begin{eqnarray}
\begin{array}{l}
\tilde m_{11}=J_H d +6tw(1-\alpha ({\bf k}))~,\\[1mm]
\tilde m_{12}=\tilde m_{21}=-J_H d ~,\\[1mm]
\tilde m_{22}=J_H d -6Jf(1-\alpha ({\bf k}))~,
\end{array}
\label{tildem}
\end{eqnarray}
with coefficients defined as
\begin{eqnarray}
\begin{array}{l}
w=<c^{\dagger}(i)c^{\alpha}(i)>~,\\[1mm]
d=2<s_z(i)S_z(i)>+<s^+(i)S^-(i)>~,\\[1mm]
f=2<S^{\alpha}_z(i)S_z(i)>+<(S^+)^{\alpha}(i)S^-(i)>~.
\end{array}
\label{wdf}
\end{eqnarray}
It is important to emphasize that $w$ is the kinetic energy and,
hence, the effective FM exchange scales with the kinetic energy.

The retarded Green's function for the boson field $B(i)$ can be
expressed as $S^B(i,j)=<R[B(i)B^{\dagger}(j)]>$. In the polar
approximation the bosonic GF can be written as
\begin{eqnarray}
S^B({\bf k},\omega)=\sum_{i=1}^{2}\frac{\tilde \sigma^i({\bf
k})}{\omega-\tilde E_i({\bf k}) +i\delta} ~, \label{gf}
\end{eqnarray}
where $\tilde \sigma^i({\bf k})$ and $\tilde E_i({\bf k}) $ are
bosonic spectral functions and the energy spectra, respectively.
$\tilde E_i({\bf k}) $ are two branches of spin excitations:
acoustic and optical  modes. In the long-wavelenght limit the
accoustic mode is $\tilde E_1({\bf k})\sim D k^2$, and the spin
stiffness scales as $D\sim (ztw-J)$ which clearly shows the
competitions of the effective,  induced by the electron hopping,
FM exchange $J_{FM}$ and AFM superexchange $J$.

There  are new internal parameters appearing in the bosonic
sector. From the analysis given above, the bosonic and fermionic
GF depend on 12 internal parameters:
\begin{eqnarray}
\begin{array}{l}
<s_z>,~ <S_z>,~ <s^+(i)S^-(i)>,~
\mu~, \\[1mm]
\Delta_{\sigma},~p_{\sigma},~ w,~ f,~ <n_{\sigma}(i)S_z(i)>~.
\end{array}
\label{ppp}
\end{eqnarray}

Now we need to construct a closed self-consistent scheme for
fixing these parameters. The static correlation functions (CF)
$C_{lm}(i,j)$ can be computed from the corresponding retarded  GF
by the fluctuation-disipation theorem:
\begin{eqnarray}
C^{F(B)}(i,j)&=& {\big (}
\begin{array}{c}
<\psi(i)\psi^{\dagger}(j)>\\
<B(i)B^{\dagger}(j)>
\end{array}
{\big )}
\\
\nonumber
 =\frac{\Omega}{(2\pi)^4}
\int d^3 k&& \!\! \int d \omega {\rm e}^{i{\bf k}({\bf r}_i-{\bf
r}_j)- i\omega (t_i-t_j)}C({\bf k},\omega)~, \label{corf}
\end{eqnarray}
where
\begin{eqnarray}
C^{F,B}({\bf k},\omega )=
\begin{array}{c}
(1+\tanh ({\omega}/{2T})) {\rm Im} S^F ({\bf k},\omega )\\[1mm]
(1+ \coth ({\omega}/{2T})) {\rm Im} S^B ({\bf k},\omega ).
\end{array}
\label{cor}
\end{eqnarray}
For convenience,  we will use the following notation for the
static CF:
 $C_{lm}=C_{lm}(i,i),~C^{\alpha}_{lm}=C_{lm}(i,i^{\alpha})$.

It should be pointed out\cite{mancini} that one can use
Eq.(\ref{corf}) for calculation of correlation functions
$C(k,\omega)$ in terms of the commutator bosonic GF only for
ergodic systems when there is  no $\delta (\omega)$ term in the
spectral density. Otherwise  one has to use the anticommutator or
causal GF for bosonic operators to take into account nonergodic
terms. In our case of FM ordered system with $<S_z\neq 0>$ and
$<s_z\neq 0>$ the bosonic GF (\ref{gf}) describes the
one-particle, spin-wave-like excitations which are ergodic for all
${\bf q}\neq 0$.

The first self-consistent equation is given by the definition of
the magnetization per site:
\begin{eqnarray}
m=\frac{1}{2}(n_{\uparrow}-n_{\downarrow})
=\frac{1}{2}[C^F_{44}-C^F_{22}]~. \label{magn}
\end{eqnarray}
The boundary condition which fixes the particle number as an
external parameter gives us the equation for fixing the chemical
potential $\mu$:
\begin{eqnarray}
n=2-C_{11}^F-C_{22}^F-C_{33}^F-C^F_{44} ~. \label{mu}
\end{eqnarray}
The other three parameters are expressed through the matrix
elements of the CF (\ref{corf}) as follows:
\begin{eqnarray}
\begin{array}{c}
\Delta _{\uparrow}=C_{11}^{F\alpha}-C_{22}^{F\alpha}\\
\Delta _{\downarrow}=C_{33}^{F\alpha}-C_{44}^{F\alpha}\\
<s^+(i)S^-(i)>=C^B_{12}~~.
\end{array}
\label{delta12}
\end{eqnarray}
The kinetic energy $w$ can also be expressed through the linear
combination of GF matrix elements:
\begin{eqnarray}
w=- (\sum_{n,m=1}^{2}+\sum_{n,m=3}^{4}) C^{F\alpha}_{nm}~.
\label{c}
\end{eqnarray}

The other  parameters are not strictly bound to the dynamics and
they must be evaluated by other means. This aspect is a general
property of the Green's function formalism \cite{mancini}. The
equations of motion are not sufficient to completely determine the
Green's functions which refer to a specific choice of the Hilbert
space. The information concerning the representation must be
supplied. As discussed in Refs.~\onlinecite{s} and
\onlinecite{mancini}, this information is supplied by determining
the parameters not bounded by the dynamics in such a way that the
Hilbert space has the right properties to conserve the relations
among matrix elements imposed by symmetry conditions. In
particular, the main attention should be paid to the fact that the
chosen represention does not violate the symmetry required by
Pauli principle. Let us note that by Pauli principle we mean all
relations among considered operators dictated by their algebra.
The Pauli principle for the electron propagator requires that
\begin{eqnarray}
\begin{array}{c}
C_{12}^{F}=0\\
C_{34}^{F}=0\\
C_{11}^{F}=C_{33}^{F}
\end{array}
~. \label{1234}
\end{eqnarray}
There is one more  self-consistent equation which connects  matrix
elements of fermionic and boson GF:
\begin{eqnarray}
C_{11}^{B}=C_{44}^{F} \label{bf}
\end{eqnarray}
To close our self-consistent scheme, we have to do some
approximation and we use the following decoupling scheme:
\begin{eqnarray}
<n_{\sigma}(i)S_z(i)>\simeq <n_{\sigma}(i)><S_z(i)> ~.
\label{decoup}
\end{eqnarray}

Having solved the system of the self-consistent equations we can
study various properties of the FKLM. In this paper we will look
at the spin wave spectrum. The study of other properties will be
published elsewhere.

In Fig.~1 we plot the dispersion of magnetic excitations along
various direction in the Brillouin zone, and we compare it with
nearest-neighbor Heisenberg form with the spin stiffness $D=
\lim_{q\rightarrow 0}\tilde E(q)/ q^{2}$, where $\tilde E(q)$ is
the acoustic mode of the bosonic spectrum calculated in the mean
field approximation (see, formula (24) in Ref.\onlinecite{6}). We
note that in this approach the acoustic spin mode already occurs
in the mean-field approximation as the result of the full
rotational symmetry of the spin system.

\begin{figure}
\epsfysize=50mm \centerline{\epsffile{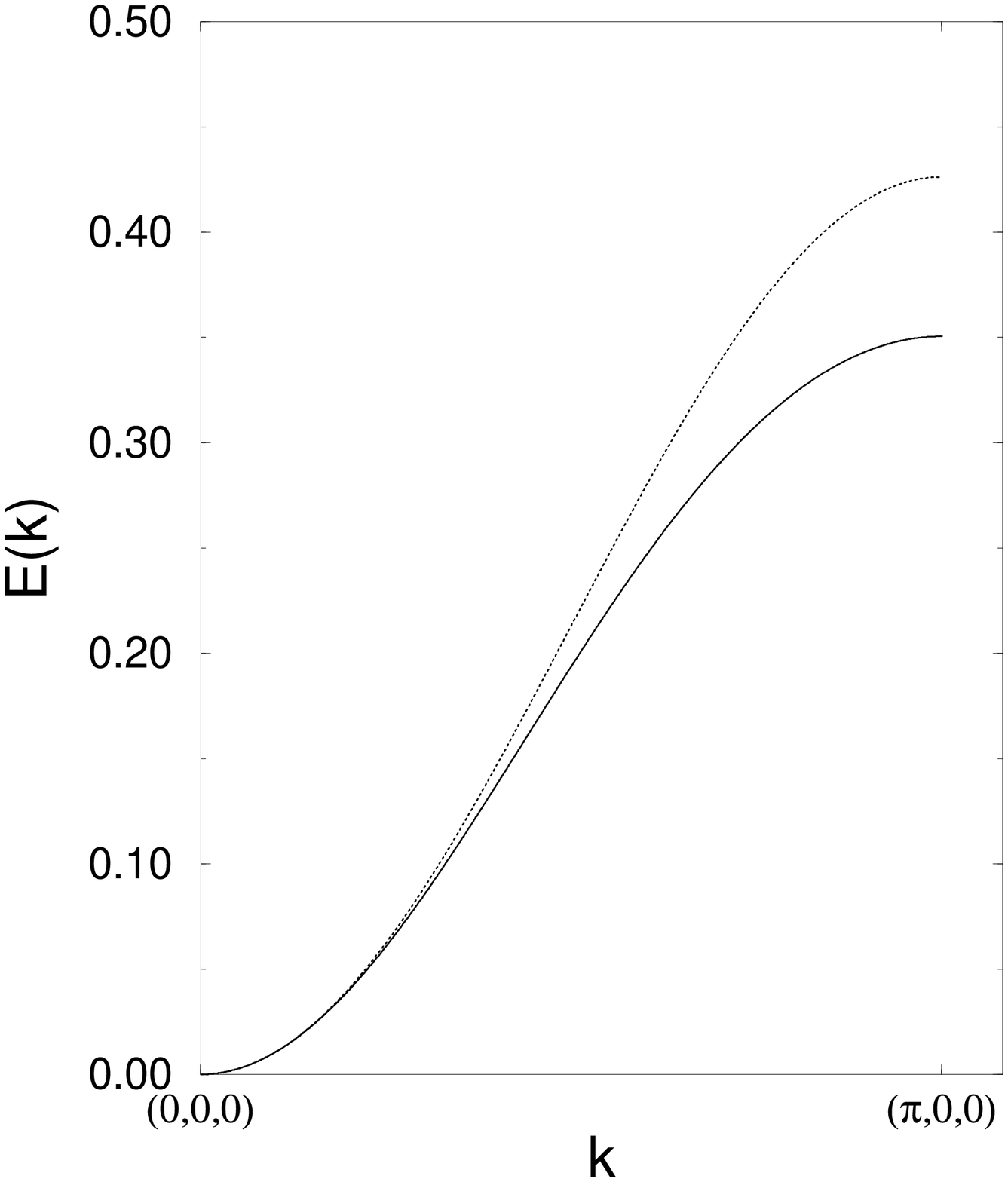}} \vspace{-0.5cm}
\epsfysize=55mm \centerline{\epsffile{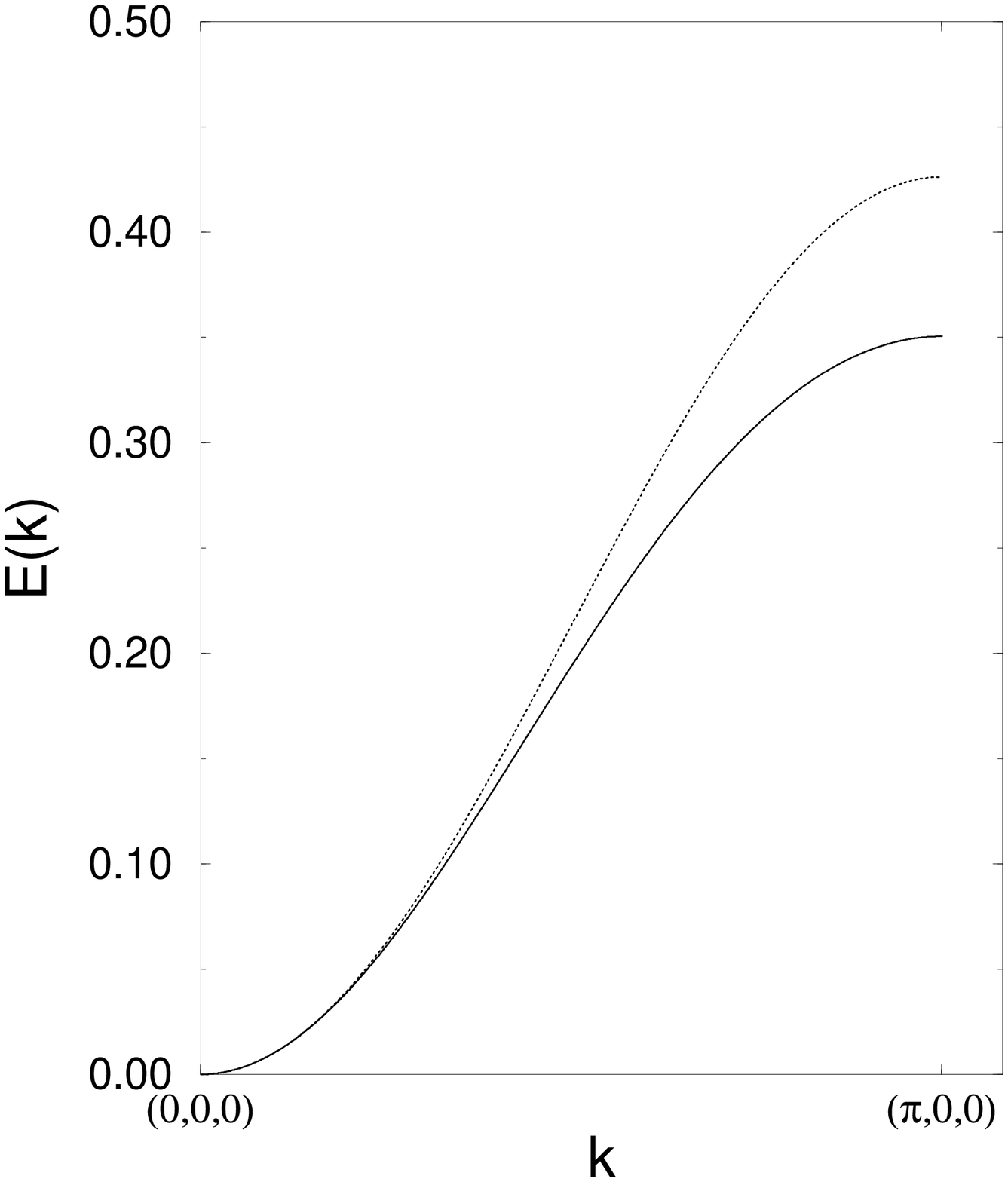}} \vspace{-0.5cm}
\epsfysize=55mm \centerline{\epsffile{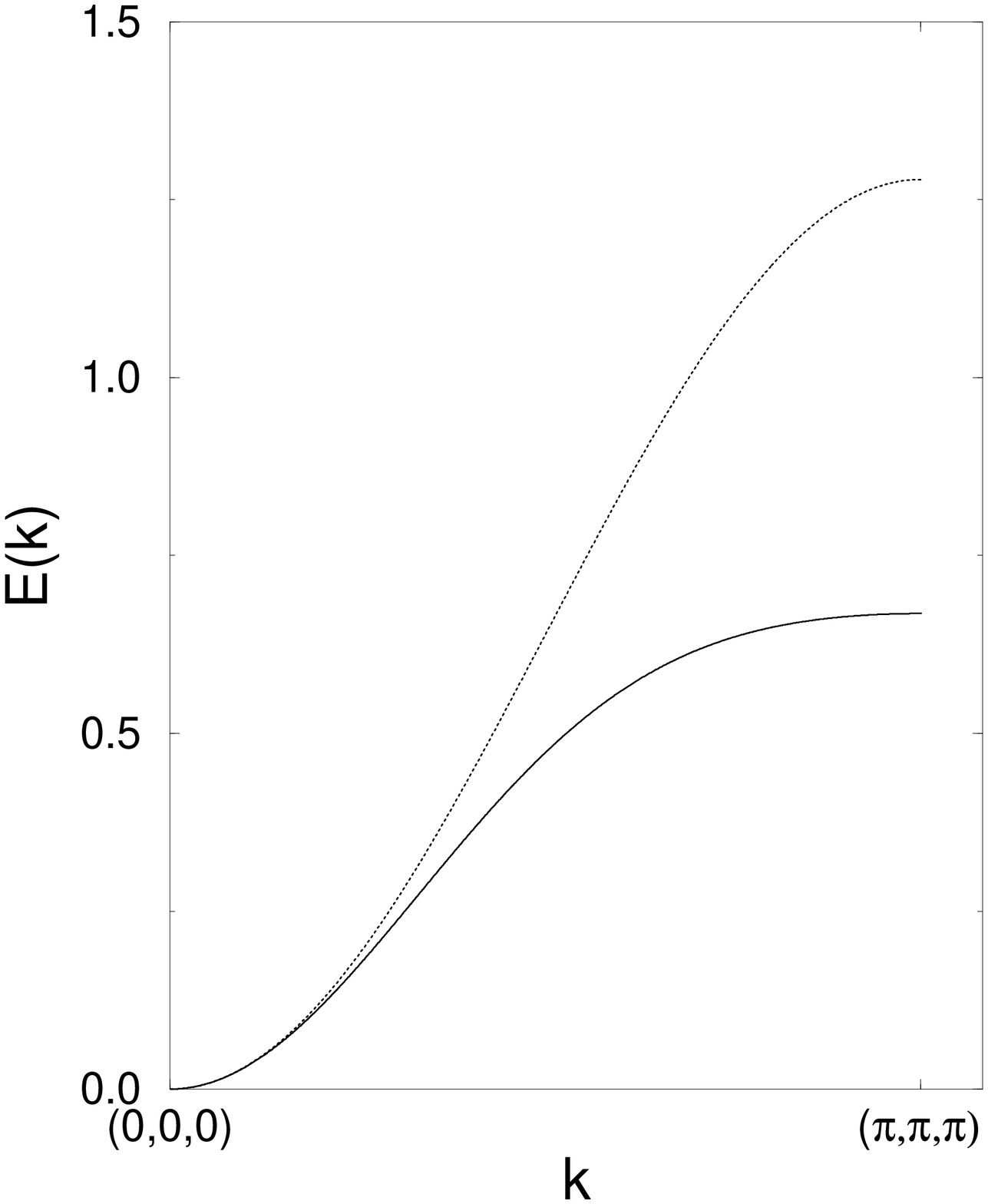}} \caption{Spin wave
dispersion obtained in FKLM along ($\pi,0,0)$, ($\pi,\pi,0$),
($\pi,\pi,\pi$) directions (solid line). The following external
parameters are chosen: $J_H=10t$, $J=0.1t$, $T=0.3t$, $U=4t$.
 Dotted line represent ideal Heisenberg spectrum with stiffness $D$.}
\label{f1}
\end{figure}

At small momenta  the spectrum exhibits conventional Heisenberg
behavior. With increasing momentum the spectrum shows the
softening which has the maximum value close to the ($\pi,\pi,
\pi$) AFM vector. However, in experiment the maximum softening was
observed in $(\pi,0,0$) and $(\pi,\pi,0)$ directions. In order to
reproduce the experimental data it is thus important to consider
orbital degree of freedom as it was considered by Khaliullin and
Kilian (see Ref.\onlinecite{4}). How strong the orbital
fluctuations modulate  exchange bonds between Mn ions depends on
their characteristic time scale. If the typical frequency of
orbital fluctuations is higher than the one of spin fluctuations,
the magnon spectrum is renormalized only due to AFM fluctuations
and then the orbital state effects the spin dynamics only by
restoring the cubic symmetry of exchange bonds. If orbitals
fluctuate slower then spins then the anisotropy imposed upon the
magnetic exchange bonds by the orbital degree of  freedom forms
spin dynamics. The presence of Jahn-Teller distortions quenches
the dynamics of orbitals and can lead to the long range orbitally
ordered state which corresponds to  the minimal frequency of
orbital fluctuations. Then both  orbital  fluctuations and the
competition between the double-exchange interaction and the AFM
superexchange interaction  are responsible for the strong
renormalization of the magnetic excitation spectrum.

In Pr$_{0.63}$Sr$_{0.37}$MnO$_3$ there is a tendency to the planar
($x^2-y^2$) ferrotype orbital ordering. That means that orbital
fluctuations mostly renormalize exchange bonds in two dimensions.
Let us note that magnons in ($\pi,\pi,\pi$) direction are sensible
to all three spatial directions of the exchange bonds and thus,
their dispersion remains unaffected by orbital fluctuations and
softening of the spectrum in this direction is purely due to the
AFM spin fluctuations. In our approach we attribute softening
phenomena to the effect of the suppression  of the  FM ordering
only by the AFM superexchange interaction and thus on the
direction ($\pi,\pi,\pi$) we estimate ordering correctly. In other
direction we have to add influence of orbital degree of freedom to
fit experimental results.

In summary, we have shown, that the spin wave spectrum obtained by
the COM analysis of FKLM shows the softening close to the magnetic
zone boundary. In our approach, we have accounted for both the
intersite AFM exchange among localized $t_{2g}$-electron spins and
the strong intra-atomic Hund coupling among the $t_{2g}$ and
$e_{g}$ electrons and proved that the competition between these
two interactions can lead to the strong renormalization of the
magnon spectrum and suppression of the FM ordering.

We thank G. Khaliullin and G. Jackeli for valuable discussions.
N.B.P. wishes to acknowledge partial support by Unit\`a I.N.F.M.
di Salerno. Financial support by INTAS Program, Grants No 97-90963
and No 97-11066, are acknowledged by N.B.P. and N.M.P..

\end{multicols}

\end{document}